\documentclass{aa}  

\usepackage{graphicx}
\usepackage{txfonts}
\usepackage{subcaption}
\usepackage[backref=page, hidelinks,colorlinks = true,citecolor = blue, linkcolor = blue, urlcolor = blue, draft=false]{hyperref}
\usepackage{etoolbox}
\usepackage{booktabs}
\usepackage{makecell}
\usepackage{upgreek}
\usepackage{bm}

\patchcmd{\chapter}{\if@openright\clearpage\else\clearpage\fi}{}{}{}

\begin{document}

    \title{3D extinction mapping of the Milky Way \break using Convolutional Neural Networks}
    \subtitle{Presentation of the method and demonstration in the Carina Arm region}

   \author{D. Cornu
          \inst{1,2}
          \and
          J. Montillaud\inst{2}
          \and
          D. J. Marshall\inst{3,4}
          \and
          A.C. Robin\inst{2}
          \and
          L. Cambrésy\inst{5}
         }

   \institute{LERMA, Observatoire de Paris, PSL research Université, CNRS, Sorbonne Université, 75104, Paris, France\\
              \email{david.cornu@observatoiredeparis.psl.eu}
         \and
            Institut UTINAM - UMR 6213 - CNRS - Univ Bourgogne Franche Comté, France, OSU THETA, 41bis avenue de l'Observatoire, 25000 Besan\c{c}on, France
         \and
          Université de Toulouse, UPS-OMP, IRAP, F-31028 Toulouse cedex 4, France
         \and
          CNRS, IRAP, 9 Av. colonel Roche, BP 44346, F-31028 Toulouse cedex 4, France
	     \and	
	     Observatoire Astronomique de Strasbourg, Université de Strasbourg, CNRS UMR 7550, 11 rue de l'université, 67000 Strasbourg, France
            }

   \date{Submitted October 14th, 2021; accepted ...}

 
  \abstract
   {Several methods have been proposed to build 3D extinction maps of the Milky Way (MW), most often based on Bayesian approaches. Although some studies {employed machine learning} (ML) methods in part of their procedure, or to specific targets, no 3D extinction map of a large volume of the MW solely based on a {Neural Network method} has been reported so far.}
   {We aim to apply deep learning as a solution to build 3D extinction maps of the MW.}
   {We built a convolutional neural network (CNN) using the CIANNA framework, and trained it with synthetic 2MASS data. We used the Besançon Galaxy model to generate mock star catalogs, and 1D Gaussian random fields to simulate the extinction profiles. From these data we computed color-magnitude diagrams (CMDs) to train the network, using the corresponding extinction profiles as targets. A forward pass with observed 2MASS CMDs provided extinction profile estimates for a grid of lines of sight.}
   {We trained our network with data simulating lines of sight in the area of the Carina spiral arm tangent and obtained a 3D extinction map for a large sector in this region ($\ell = 257\!-\!303$ deg, $|b| \le 5$ deg), with distance and angular resolutions of 100 pc and 30\arcmin, respectively, and reaching up to $\sim\!10$ kpc. Although each line of sight is computed independently {in the forward phase}, the so-called fingers-of-God artifacts are weaker than in many other 3D extinction maps. For the training phase, we found that our CNN was efficient in taking advantage of redundancy across lines of sight, enabling us to train it with only 9 lines of sight simultaneously to build the whole map.}
   {We found deep learning to be a reliable approach to produce 3D extinction maps from large surveys. 
   {With this methodology, we expect} to easily combine heterogeneous surveys without cross-matching, and therefore to exploit several surveys in a complementary fashion.}

   \keywords{Methods: statistical - ISM: dust, extinction -
            ISM: structure - Galaxy: structure - Galaxy: local interstellar matter
            }

\maketitle
%

\section{Introduction} 
Unveiling the structure of the Milky Way Galaxy is a challenging task due to our location within the Galactic disk. Since the founding work by \citet{marshall_modelling_2006}, a number of methods have been proposed to characterize the 3D distribution of interstellar matter traced by interstellar extinction. In such studies, the observed stellar radiation is compared to the expected stellar radiation either from individual stars or, statistically, from stellar populations. The discrepancy between them is interpreted as the result of interstellar extinction, and the knowledge on stellar distances is used to constrain the 3D distribution of extinction.
\citet{green_3D_2019} derived the 3D dust extinction from the combination of Gaia DR2 parallaxes \citep{Gaia_Collaboration_2018_global}, Pan-STARRS 1 \citep{Chambers_2016} and 2MASS \citep{skrutskie_two_2006} photometry using a Bayesian approach implemented with Markov chain Monte-Carlo (MCMC), covering typically the first two to three kiloparsecs with an angular (multi-)resolution between $\sim\!3\arcmin$ and $55\arcmin$. In spite of an effort to explicitly take the correlations between adjacent sightlines into account, the map still includes the so-called fingers-of-God artifacts (i.e. spreading of extinction estimates along the line of sight due to distance uncertainties). In contrast, \citet{lallement_gaia-2mass_2019} used a hierarchical Bayesian inversion algorithm applied to a star catalog compiled from a cross-match of Gaia and 2MASS data. They obtained a high distance-resolution map in a 3-kpc range, essentially devoid of fingers-of-God artifacts.
In contrast to \citet{marshall_modelling_2006}, based solely on 2MASS data, these studies \citep[see also][]{sale_3d_2014,chen_three-dimensional_2019,leike_resolving_2020} take advantage of multiple surveys (or plan to do so) to benefit from a better characterization of the star intrinsic emission or distance. However, even when near infrared (NIR) data were used, the proposed methods required to cross-match the NIR and visible catalogs, so that the final map inherited the limitations from both spectral domains. The maps resulting from such methods tend to be limited to the first few kpc from the Sun because interstellar extinction is higher in the visible range.

    \begin{figure*}[t]
        \centering
        \includegraphics[width=0.65\hsize]{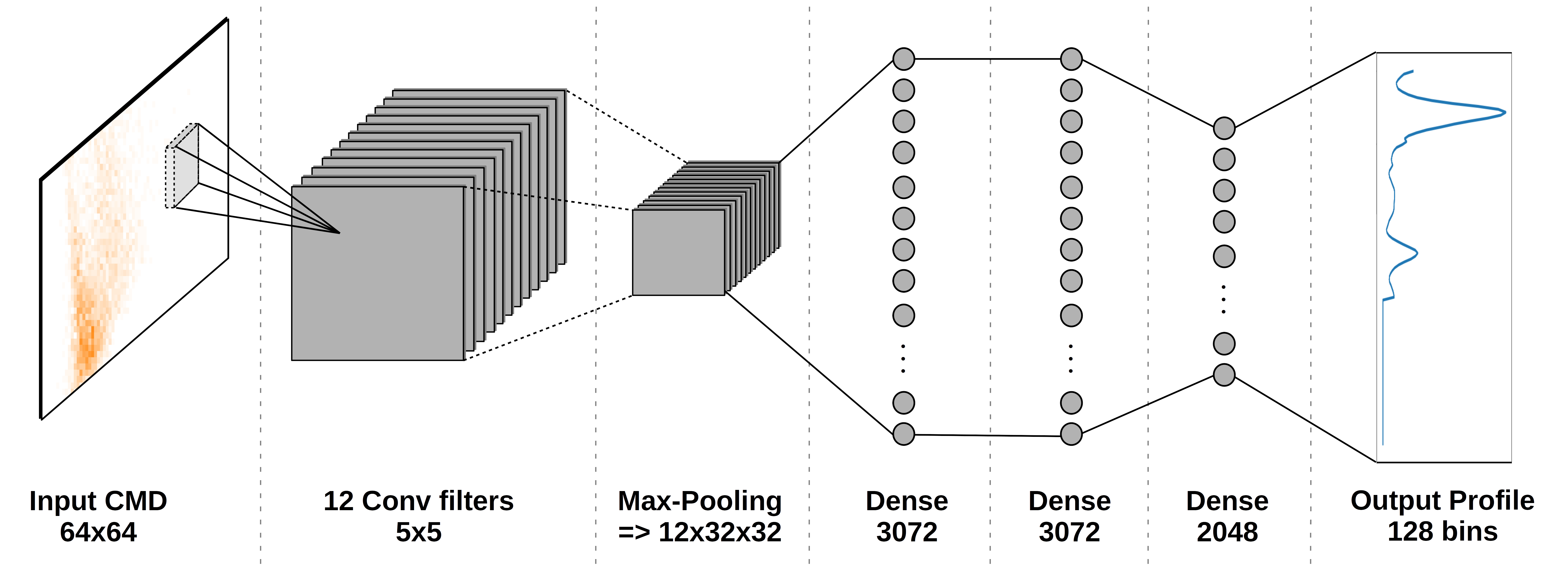}
        \caption{CNN architecture that provided the best performance for extinction map reconstruction}
        \label{cnn_arch}
    \end{figure*}
    
    \begin{figure}[t]
        \centering
        \includegraphics[width=1.0\hsize]{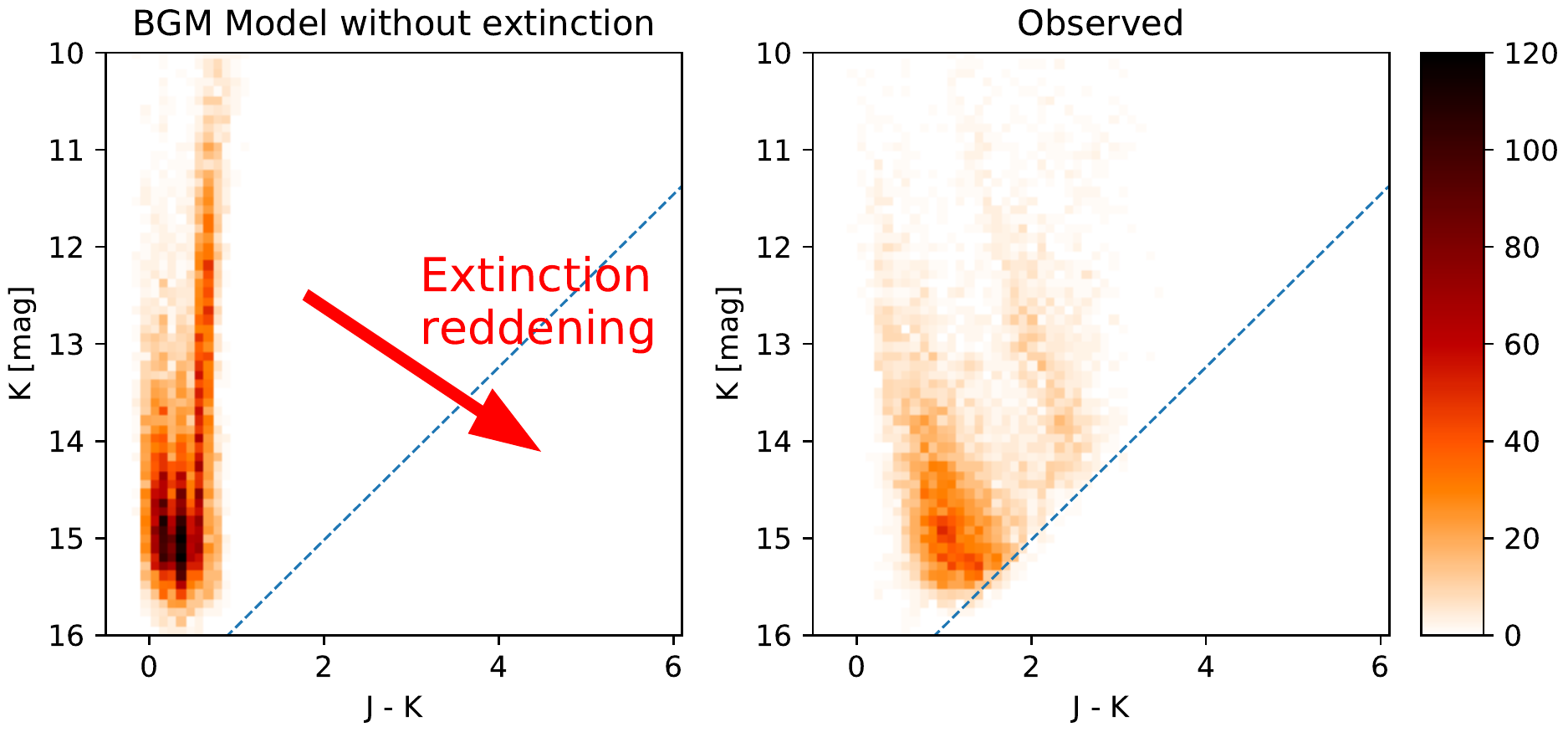}
        \caption{2MASS [J-K]-[K] CMD comparison between model prediction without extinction on the {\bf left} and observed extinction on the {\bf right} for the same sightline. The two diagrams are obtained from a 0.25 degree radius centered at galactic coordinates $\ell = 280$ deg, $b = 0$ deg. The dashed blue line corresponds to our cut on observed magnitudes and colors.}
        \label{ext_cmd_effect}
    \end{figure}

In this letter, we propose a new strategy, based on convolutional neural networks (CNN). Ultimately, we expect such approaches to make it easy to combine different stellar catalogs without cross-matching them, and therefore to take advantage simultaneously of the Gaia parallaxes and of the greater reach of NIR surveys, or of the spectral type classification from PAN-STARRS 1 data. We report here a new approach using CNNs, based on the comparison between 2MASS data and mock data synthesized with the Besançon Galaxy Model \citep[BGM, ][]{robin_synthetic_2003, lagarde_three-dimensional_2014}. We present the method in the next section. The obtained 3D extinction maps are presented in Sect.~\ref{sec:results}, and the implications for future studies based on multiple surveys are discussed in Sect.~\ref{sec:discussion}.

\section{Method}\label{sec:method}

Our method is based on the supervised training of CNNs \citep{Lecun-95, Lecun-2015}. We built the training sample as follows.
We used the BGM\footnote{Version mev1802} to produce statistically-representative star lists based on cone shaped sightlines, from which we used solely the positions and the mock 2MASS magnitudes, namely the J~(1.235\,$\rm{\upmu m}$), H~(1.662\,$\rm{\upmu m}$), K~(2.159\,$\rm{\upmu m}$) bands. From the selected bands it is possible to build [J-K]-[K] CMDs for different sightlines that are sensitive to both the reddening and the brightness decrease induced by extinction. Our star lists were produced without extinction; we applied our own mock extinction profiles to produce the corresponding extincted CMD following the extinction law by \citet{cardelli_relationship_1989}. The induced movement and deformation of the star distribution in this 2D histogram can be used to reconstruct the corresponding extinction profile. To produce realistic profiles we used Gaussian Random Fields \citep[GRFs,][]{coles_lognormal_1991, Sale_2014} composed of two contributions to represent both large (>2~kpc) and small (<2~kpc) distance scales of the interstellar matter distribution. Finally, to avoid confusing the CNN with information which would be impossible to reconstruct, (i) we set to zero the extinction profiles beyond the distance at which there are fewer than 50 detected stars further away in the list, and (ii) we clipped the maximum value of the extinction profiles to 50~mag/kpc in the V band. On these simulated stars, magnitude cuts (Fig.~\ref{ext_cmd_effect}) and photometric uncertainties were taken into account to match realistic 2MASS observations. These cuts and uncertainties were realized statistically for each CMD using a selection function calibrated on a 1 square degree region around the central region $\ell = 280$ deg, $b = 0$ deg.

The method consists in training a CNN to perform the inverse transformation, from an extincted CMD to the associated extinction profile. For each sightline, we sampled the CMD on $64 \times 64$ bins in the ranges $-0.5<$~[J-K]~$<6.1$ and $10<$~[K]~$<16$. This defines the dimension of the network input layer. The inner network architecture is fairly shallow as illustrated in Fig.~\ref{cnn_arch}, with all neurons being ReLU activated \citep{he_delving_2015}. Despite extensive testing of larger and deeper architectures, the one presented here performed the best. This is likely because the amount of extinction at a given distance is essentially encoded in the displacement of stars in the CMD (Fig.~\ref{ext_cmd_effect}). In other words, it involves a spatial information in the CNN input layer, which is more suitably interpreted by fully connected (FC, or dense) layers. Still, the few FC layers need to be large to achieve good results, indicating that the relation that we seek to construct remains of high complexity. The dropout technique \citep{srivastava_dropout_2014, gal_dropout_2015} is applied to the first two dense layers composed of 3072 neurons each, with a 10\% dropout rate. With this technique, only 90\% of the neurons are randomly selected and used at a time. For inference this is repeated 100 times for each input with a different dropout selection. This acts as a Monte Carlo (MC) method to construct probability distributions, instead of single value predictions. The output layer is the reconstructed extinction profile sampled using 128 bins of 100 pc each with a direct linear activation for each bin.

    \begin{figure*}
        \centering
        \begin{subfigure}[t]{0.75\textwidth}
        \includegraphics[width=1.0\hsize]{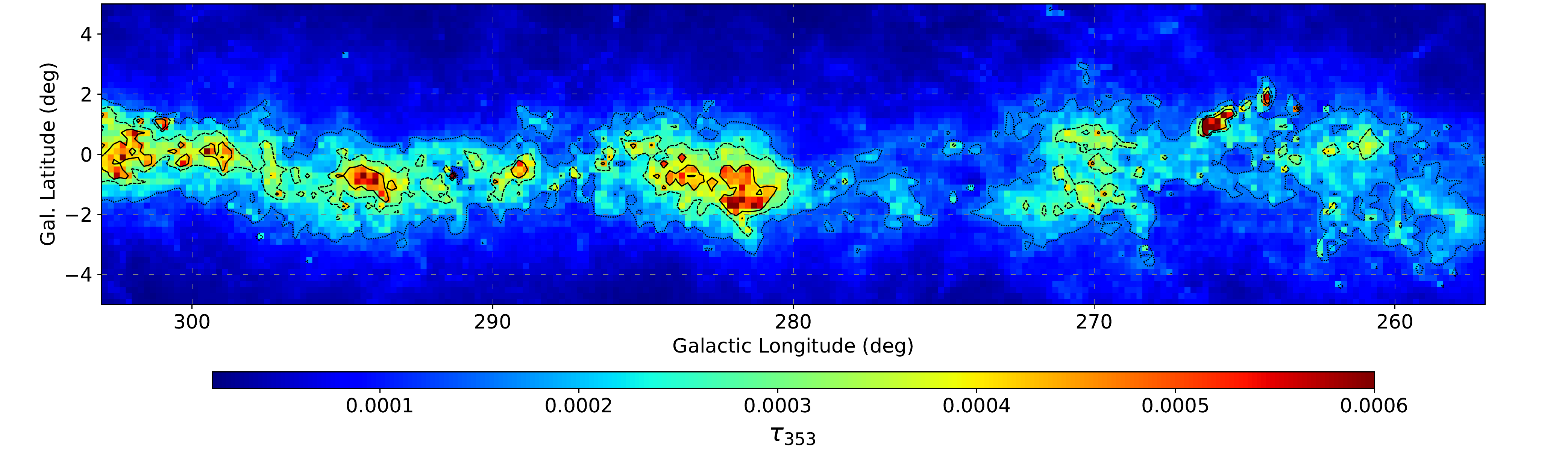}
        \end{subfigure}
        \begin{subfigure}[t]{0.75\textwidth}
        \includegraphics[width=1.0\hsize]{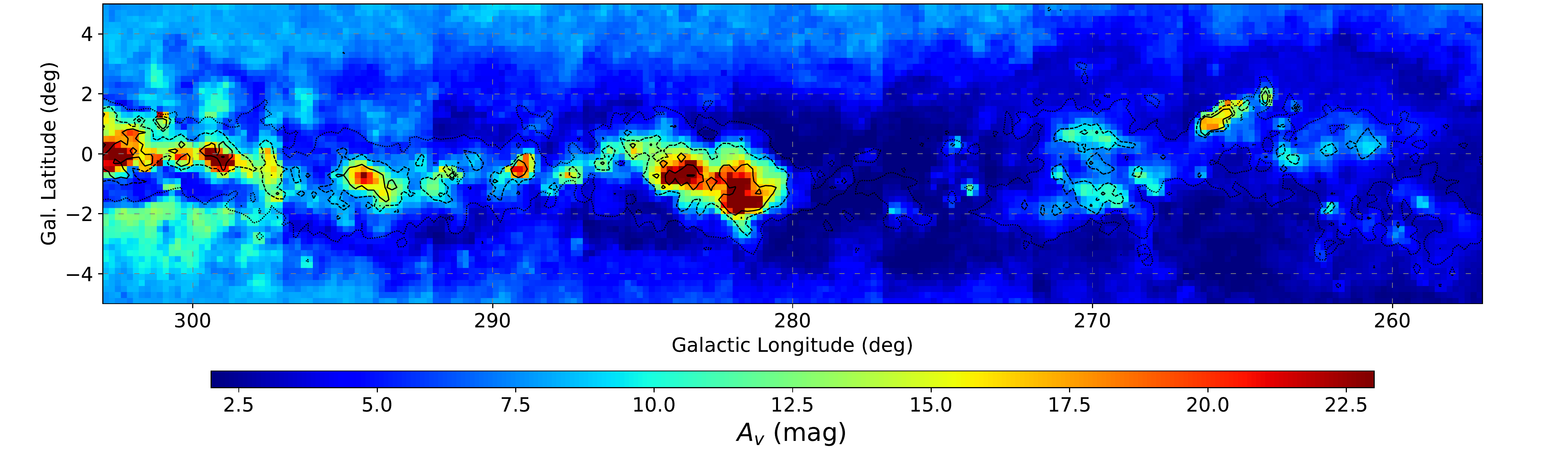}
        \end{subfigure}
        \caption{Comparison of our integrated 3D extinction map (bottom) and the Planck dust opacity at 353 GHz (top) in the Carina-arm region. In both frames the contours are Planck levels at $\tau_{353}=0.00016$, 0.00028, and 0.0004.}
        \label{fig:integ_maps}
    \end{figure*}

Considering that a simulated non-extincted CMD represents a specific sightline, a network trained to reconstruct profiles on a given CMD cannot be efficiently applied to another sightline. To construct a network that can be applied to a larger range of galactic positions, we added a second channel to the input layer that contains a reference non-extincted CMD from the model. Using this input construction, the network learns the difference between the reference and an extincted CMD, allowing it to be trained simultaneously for several sightlines and to interpolate information between them. During the training phase all (extincted and non-extincted) CMDs come from the BGM using mock extinction profiles, while during the inference phase the extincted CMDs are replace by observed data for which we want to reconstruct the profiles, but the reference CMDs remain provided by the BGM for each sightline. In practice, it is very computationally intensive to generate a high resolution grid of BGM reference sightlines. For this reason, we keep the same reference CMD for a few degree wide sky region.

Training the network on a single sightline usually requires up to $5\times 10^5$ pairs of mock profiles and CMDs, with around 5\% of the examples kept as validation dataset and 1\% as test dataset. However, when training for multiple reference sightlines distributed every 5~degrees, the amount of examples per reference can be contained down to $2\times 10^5$, with a typical global dataset size of $1.8\times10^6$ examples (which requires up to 120 GB of RAM). This confirms that part of the information can efficiently be shared between reference CMDs, and that training a single network on multiple sightlines is more efficient than training different networks for several generalization regions. We trained our network using our custom CNN framework \href{https://github.com/Deyht/CIANNA}{CIANNA}\footnote{CIANNA is open source and freely accessible through GitHub https://github.com/Deyht/CIANNA. The version used corresponds to commit a6b903c.} (Convolutional Interactive Artificial Neural Networks by/for Astrophysicists) that handles GPU acceleration allowing to train in a few hours on a Nvidia Tesla V100 GPU. For inference, we used MC model averaging with 100 realizations for each sightline (Fig.~\ref{fig:prediction_test}) and used the averaged value to produce the maps presented in Sect.~\ref{sec:results} along with prediction dispersion maps as illustrated in Fig.~\ref{fig_std_polar_map}.

\section{Results}\label{sec:results}

\begin{figure*}
    	\centering
    	\begin{subfigure}{0.4\textwidth}
    	\includegraphics[width=1.0\hsize]{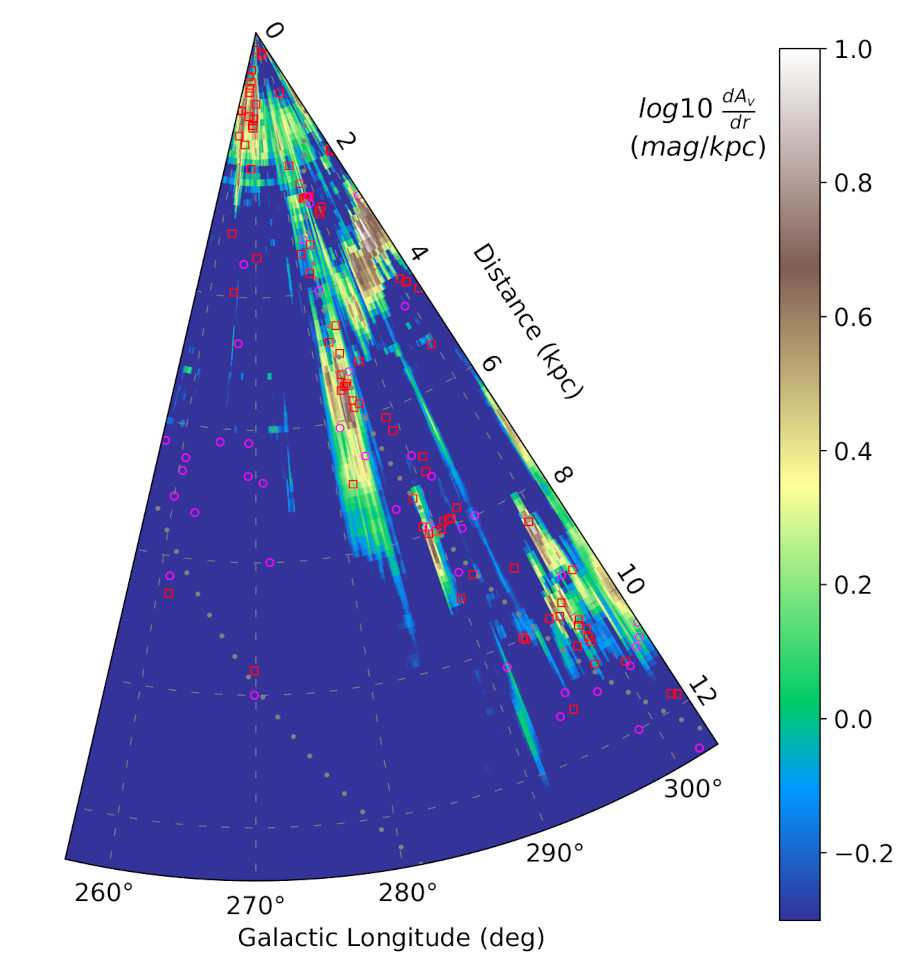}
    	\end{subfigure}\hspace{0.5cm}
    	\begin{subfigure}{0.4\textwidth}
    	\includegraphics[width=1.0\hsize]{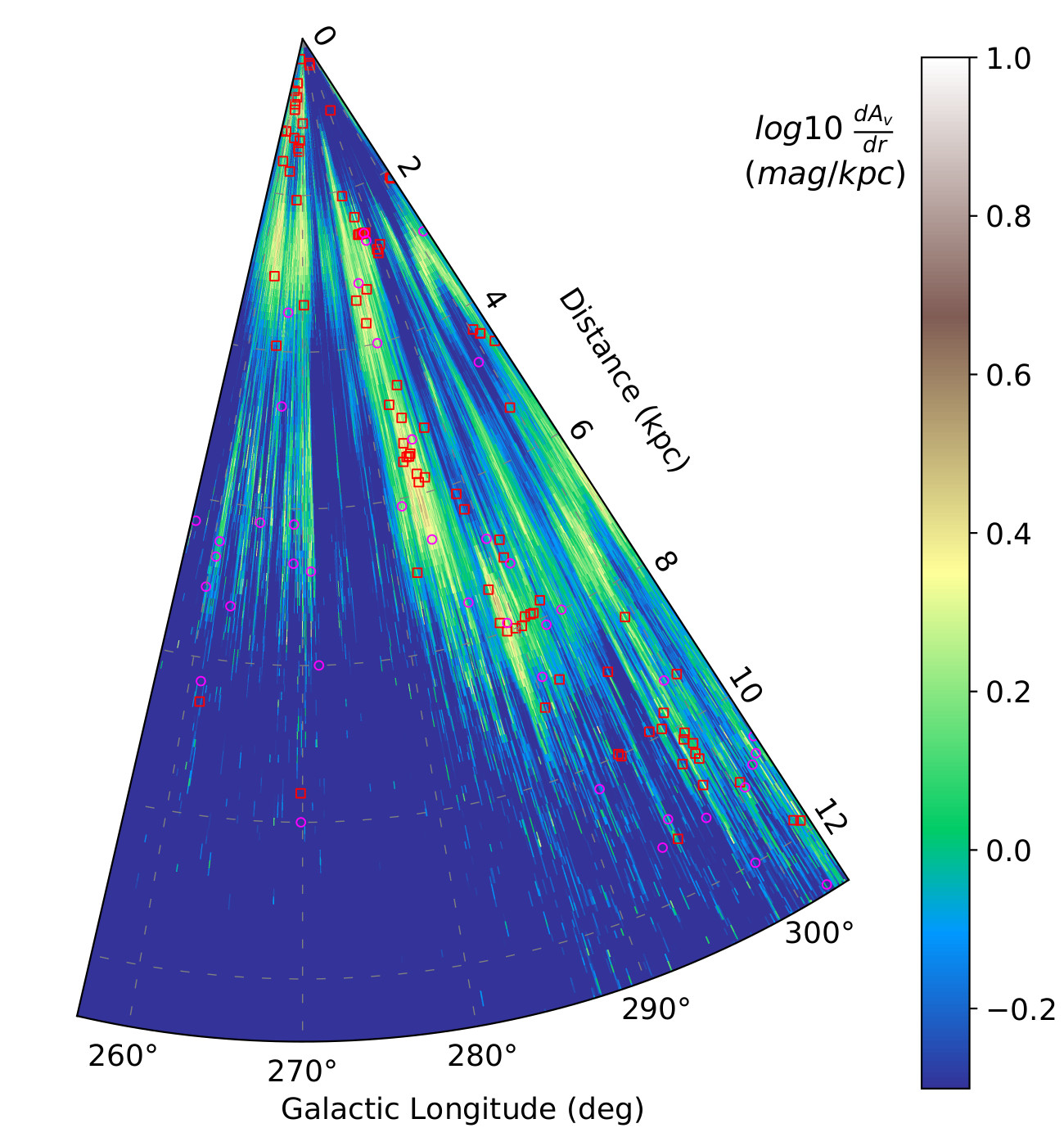}
    	\end{subfigure}
    	\caption{Face-on view of the Galactic Plane for $|b|<1$\,deg in polar coordinates in the Carina-arm region. The left and right frames show our prediction and the prediction by Marshall et al. (in prep.), respectively. The red open squares and purple open circles show the HII regions and giant molecular clouds, respectively, as tabulated by \citet{hou_observed_2014}. In the left frame, the gray dots show the simple spiral arm model from \citet{marshall_modelling_2006}.}
    	\label{multi_los_2mass_polar_sky}
    \end{figure*}

We used the method presented in the previous section to construct a 3D extinction map of the Carina arm region, i.e. for $\ell$ between 257 and 303 degrees, and $|b|<5$ degrees. We tiled this region with nine rectangular zones centered on nine reference sightlines located at $\ell = 260, 265, \dots, 300$ deg, and $b=0$ deg. The region is sampled by pixels separated by 12 arcmin in the plane of the sky, and the extinction in each pixel is computed from the 2MASS sources within a cone of 15 arcmin radius. This ensures that the Nyquist criterion is fulfilled.

We emphasize that there is no forced constrain on the total integrated extinction of a given sightline in the network. Each neuron in the output layer is only penalized in order to reconstruct the local extinction value, still the complete network is free to evaluate that the summed value of all neurons is a valuable information. Additionally, while there is a partial overlap between two adjacent sightlines, there is no forced coherence between them. The only information that is shared between sighltines is the reference CMD inside a given reference interval but they are considered as independent by the network construction.

Figure~\ref{fig:integ_maps} shows the obtained extinction map, integrated along the line of sight, and compared with the Planck opacity map at 353 GHz. This Planck map is considered to be a good proxy for the distribution of interstellar matter, although it is known to be biased by dust temperature gradients, especially in the Galactic Plane (GP) and in regions irradiated by young stars \citep{planck_collaboration_xi_planck_2014}. The general agreement between the two maps is clear for the structures at large opacity, implying that our method captures the essential of the dense interstellar medium. Differences can still be spotted in the relative levels of the structures. For example, we find a larger extinction in the major structure at $\ell\!\sim\!285$\,deg than in the compact clump at $\ell\!\sim\!266$\,deg, while the reverse is observed in the Planck map. More discrepancies are found at intermediate opacity ($\sim\!2 \times 10^{-4}$), which tends to be systematically underestimated in our map.

Two types of artifacts can also be noticed. The tiling effect at the junction between the nine zones arises from the sudden change in reference sightline. The other artifacts are visible at larger galactic latitudes, where we find an increase in extinction instead of the expected decrease. It arises from the decrease in star count when $|b|$ increases. Since the reference sightline is at $|b|=0$, the difference in star count is interpreted by the network as a signature of extinction. In addition, the artifacts present a dependence in longitude which is particularly visible in the tile centered at $\ell = 300$\,deg. These effects could be mitigated by splitting each zone into several zones along the longitude and latitude axes, at the cost of a heavier training. This artifact is also very likely to be linked to the fact that we generalized our magnitude cuts and uncertainties from the $l=280$ deg, $b=0$ deg sightline over the all map, which increase discrepancies between the observed and modeled CMD. Having these selection function calibrated for each reference sightline would certainly also help reduce the artifact. For the present paper, it is sufficient to conclude that our extinction map is only marginally affected by this artifact in the range $|b|<2$ deg.

Figure~\ref{multi_los_2mass_polar_sky} (left) shows a face-on view of the GP in polar coordinates, obtained by integrating our 3D map over the $|b|<1$\,deg range. Structures are detected between 1 and 10\,kpc along the Carina arm. A particularly good agreement is found with the distribution of HII regions and giant molecular clouds (GMCs) tabulated by \citet{hou_observed_2014} for the arm tangent near $\ell = 280$\,deg and $d=6$\,kpc. The agreement is less clear beyond the tangent, around $\ell = 290$\,deg and $d=8$\,kpc, where part of the structures seem to be disturbed by extinction at shorter distances ($d\!\sim\!4$\,kpc) which are likely to be artifacts of the method. Interestingly, a notable amount of extinction is found at large distances ($d\!\sim\!10$\,kpc) for $\ell \sim\!295$\,deg, in good agreement with a cluster of HII regions and GMCs, suggesting that this is a genuine detection. In the $\ell=255\!-\!275$\,deg region, the Vela complex is detected near 2\,kpc (better seen in Cartesian coordinates, Fig.~\ref{fig:cartesian_map}), but beyond this distance the map is devoid of extinction in spite of numerous known GMCs. This reflects the difficulty to detect faint structures, especially for sightlines with lower star counts, when looking closer to the Galactic anti-center.

These results are contrasted with those from the new map by Marshall et al. (in prep., Fig.\ref{multi_los_2mass_polar_sky} right), also obtained from the comparison between 2MASS data and BGM simulations, with a MCMC approach, and which can be considered as an update of the map by \citet{marshall_modelling_2006}. From this comparison, our CNN-based method appears to be more sensitive to large distance structures, but more prone to major artifacts (e.g. the structure at $\ell=290$\,deg and $d=4$\,kpc). In a large part of the map ($257<\ell<280$\,deg and $2<d<13$\,kpc) our method fails to detect faint extinction where other maps predict noisy structures. 
This is certainly due to a lack of remaining visible stars that represents these distances in the corresponding extincted CMDs. Due to the star count threshold introduced in our training profile construction, the network interprets such regions as a zero profile after a given distance (Sect.~\ref{sec:method} and Fig.~\ref{fig:prediction_test}). The lack of remaining stars in this region is the combination of (1) a drop in total star count par sightline for this range of galactic longitude, and (2) the result of the strong foreground extinction corresponding to the Vela complex.
In the rest of the map, our method produces cleaner maps with less striking fingers-of-God artifact and more coherent structures in dense regions. Interestingly, the results obtained on the validation set show that with a sufficient number of stars, the method can retrieve faint and diffuse extinction even between two important extinction peaks. For this reason, it is likely that the very clean region about $l=295$\,deg, $d=7$\,kpc represents a significant non-detection of extinction between two dense structures. For small distances (< 3 kpc), the structures found in the new Marshal map resemble those present in the map by \citet{lallement_gaia-2mass_2019}, although the latter study finds them to be closer. In contrast, our method tends to merge several of those structures, similarly to the case illustrated in Fig.~\ref{fig:prediction_test}(middle row), and to predict them at a closer distance than the Marshall et al. (in prep.) map.

\section{Discussion and conclusion}\label{sec:discussion}

The methods employed so far to infer 3D extinction were mostly based on a variety of Bayesian approaches, like MCMC \citep[e.g.]{green_3D_2019,sale_3d_2014}, inversion techniques \citep{lallement_gaia-2mass_2019}, variational inference \citep{leike_resolving_2020}, or Gaussian processes \citep{rezaei_kh._inferring_2017}. Although some of these methods are sometimes tagged as machine learning (ML) approaches, more widely adopted methods of ML were only scarcely used for 3D extinction maps. A genetic algorithm was experimented by \citet{marshall_distribution_2009} but only for selected target clouds, and \citet{chen_three-dimensional_2019} used random forest to determine the color excess toward each star, but then a Bayesian method similar to \citet{green_3D_2019} was implemented to build the 3D map. The present work is the first, to our knowledge, to experiment neural networks to directly build a 3D extinction map.

Our CNN-based method turns out to be capable of reconstructing a realistic morphological distribution of interstellar matter in the plane of the sky through integrated extinction, although it was not constrained to reproduce this quantity. It has demonstrated the capability to efficiently reproduce dust extinction profiles for individual sightlines and is a suitable tool to produce large-scale 3D Galactic extinction maps in the GP.

By construction, the method is less prone to the usual fingers-of-God artifacts than observed in similar maps. Structure elongation in distance is contained by both the choice of the typical scales in mock profiles of the training sample and by the effect of smoothing the CMDs by convolution in the first network layer. It is also notable that structures tend to be more coherent between adjacent sightlines than with the MCMC method, producing smooth features even in the plane of the sky. Importantly, we did not force this coherence since distance bins correspond to individual neurons and extincted input CMDs are different for adjacent sightlines. Still, the coherence is encouraged by the use of the same reference CMD in a given zone, but also by the global training over several reference sightlines, allowing the network to better infer the general properties. The method is able to detect compact structures at large distances that are coherent with known HII and GMC regions and which were missed in similar maps.

However, our approach is hampered by its strong sensitivity to the realism of the training dataset. Subtle differences between observed data and modeled training data may result in large artifacts. These differences can come from the building of the extincted CMDs from the mock stars, since each step (Sect.~\ref{sec:method}) comes with its own inaccuracies. The BGM can also contribute to the differences between observed and modeled CMDs, for example because it is not designed to describe out-of-equilibrium dynamical structures or small scale structures (e.g. open clusters). In addition, the use of the same reference over a large sky area implies that the modeled data may not accurately match the observed data. This effect is minor when looking toward the outer Galaxy, but it becomes more stringent when getting closer to the Galactic center, which can result in major artifacts. While it would be very difficult to have a CMD for each sightline, a promising development consists in constructing an adaptive reference grid that better samples the sky where CMDs change quickly with sky coordinates. This would be particularly needed when approaching the Galactic center, and to account for variations in Galactic latitude.

Our tests showed that a critical parameter in our method is the limit we imposed, when building the training sample, on the minimum number of background stars required to derive an extinction estimate. This threshold makes it difficult to apply the method in regions with very low stellar number densities. Interestingly, the strategy combining multiple reference sightlines enabled us to decrease this threshold value by a factor of two compared to a single reference training. Therefore, a useful future development of our method would be to introduce an adaptive star count threshold.

Finally, an important strength of the CNN approach for future improvements is the possibility to combine information from various surveys in a very efficient way, without cross-match. Other statistical 2D diagrams of various natures can easily be added in the form of input layers, along with the corresponding modeled references when necessary. From this heterogeneous data combination, the Neural Network structure would learn to weight and combine the information of different nature depending on the context. An example could be to add a Gaia [$G_{\rm bp} - G_{\rm rp}$]-[$\varpi$] or [$G_{\rm bp} - G_{\rm rp}$]-[$G$] diagram. This would add information for short distances and still conserve the large distance information from 2MASS, even for stars too extincted to be detected by Gaia. This approach could theoretically be pushed to add an arbitrary number of surveys that contains information about extinction.

\begin{acknowledgements}
    This research was supported by the Programme National “Physique et Chimie du Milieu Interstellaire” (PCMI) of CNRS/INSU with INC/INP co-funded by CEA and CNES as part of the GALETTE project.
    We also acknowledge the International Space Science Institute, Bern, Switzerland for providing financial support and meeting facilities.
    Training computations have been performed on the supercomputer facilities of the "Mésocentre de calcul de Franche-Comté". Inference computations have been performed on the computing server of the CompuPhys Master program, with the kind help of the "Plateforme d'informatique scientifique d'UTINAM" (PISU) staff.
    This publication makes use of data products from the Two Micron All Sky Survey, which is a joint project of the University of Massachusetts and the Infrared Processing and Analysis Center/California Institute of Technology, funded by the National Aeronautics and Space Administration and the National Science Foundation.
\end{acknowledgements}

\bibliographystyle{aa}
\bibliography{cnn_extinction}

\begin{appendix}

\begin{figure*}
    	\centering
    	\renewcommand{\thefigure}{A.1}
    	\includegraphics[width=0.85\hsize]{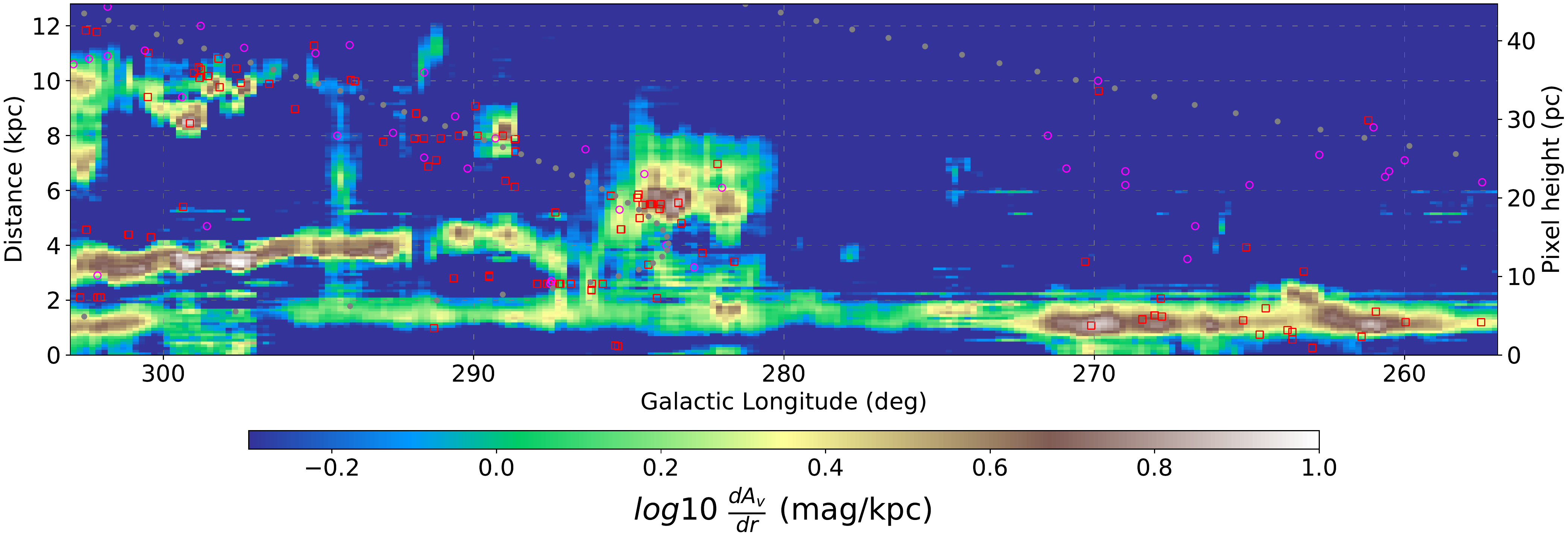}
    	\caption{Face-on view of the Galactic Plane $|b| < 1$\,deg in Cartesian galactic-longitude distance coordinates in the Carina arm region. The axis on the right border corresponds to the pixel vertical size as a function of the distance induced by the conic shape of the sightline. The symbols are the same as in Fig.~\ref{multi_los_2mass_polar_sky}.}
    	\label{fig:cartesian_map}
\end{figure*}

\begin{figure*}
        \centering
        \renewcommand{\thefigure}{A.2}
    	\includegraphics[width=0.95\hsize]{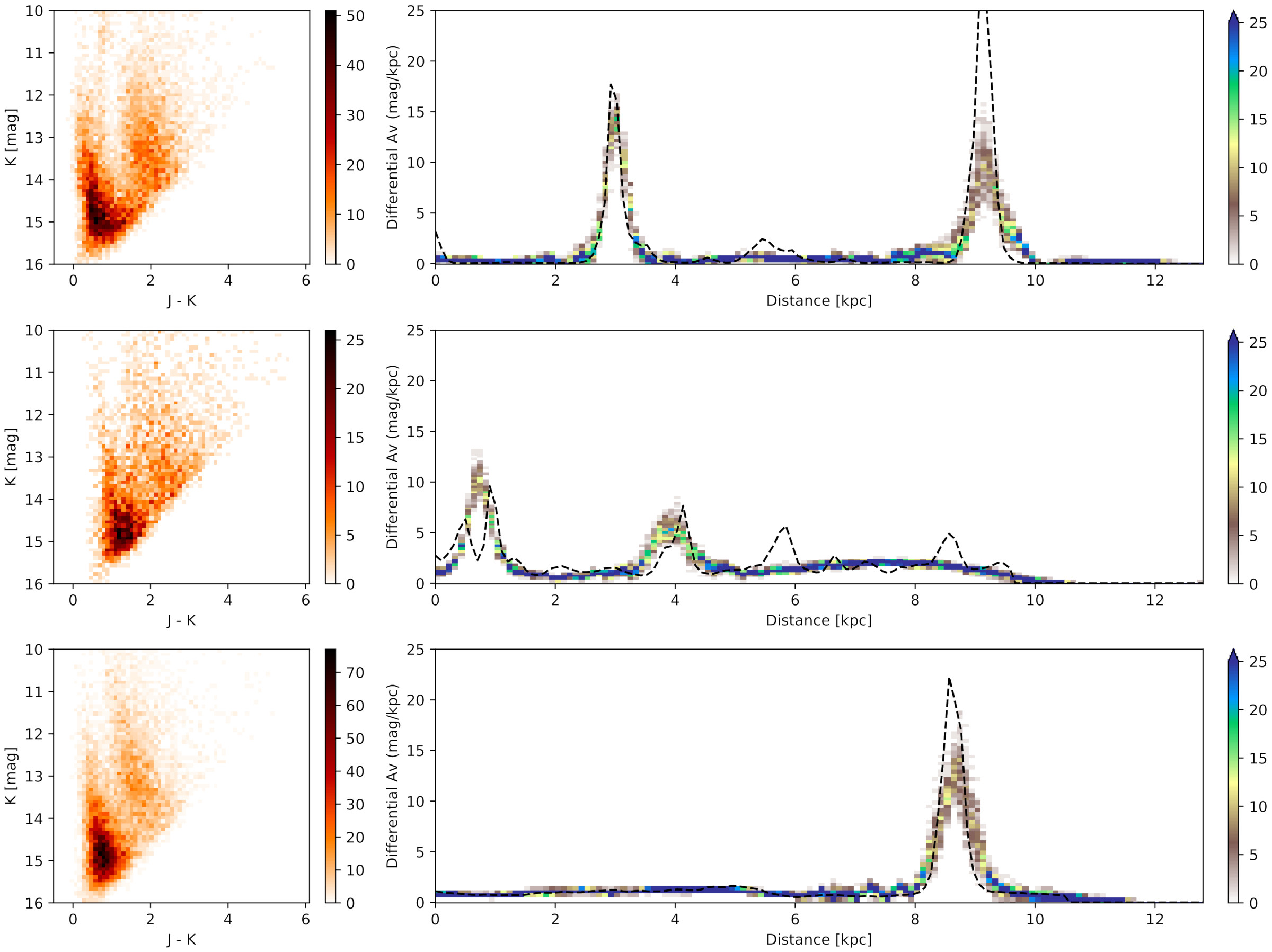}
    	\caption{Predicted and target profile comparison from the test dataset. These examples are for the $\ell=280$ and $b=0$ reference sightline. For each case the left frame shows the input extincted CMD and the right frame presents the extinction profiles. The dashed black line is the target profile. The network prediction is presented in the form of a color-coded 2D histogram for each distance bin, representing the probability distribution based on 100 MC model dropout predictions.}
    	\label{fig:prediction_test}
\end{figure*}

\clearpage

\section{Additional materials}

Figure~\ref{fig:cartesian_map} is a Cartesian representation of same result as the left frame of Fig.~\ref{multi_los_2mass_polar_sky}. This view better represents how the network interprets the individual sightlines and constructs the map, with no assumptions on pixel size. The cone-shaped dilution effect is accounted for during the application of the profile to the star list, but this is not a direct information given to the network. This illustration provides a better view of the nearby structures and their connections, at the cost of physical size deformation as a function of distance. It also allows to more accurately identify structure positions and help to identify those that align on the same sightline.

Figure~\ref{fig:prediction_test} presents a few comparisons of predicted profiles with the associated target profiles along with the corresponding extincted CMDs. Thanks to dropout, 100 MC model predictions are obtained and used to build a discretized probability density distribution of differential extinction as a function of distance. This is represented by a 2D histogram and helps evaluating the dispersion in the CNN predictions.
The figure demonstrates that the network is able to reconstruct extinction peaks at different distance ranges and that it successfully detects secondary structures. It also shows that the predicted value of the extinction maximum is well reconstructed for primary peaks, and remains sufficiently satisfying in the secondary structures to be confident in the predicted morphology. The second row illustrates the tendency of the method to smooth structures closer than 500 pc.

\begin{figure}
        \centering
        \renewcommand{\thefigure}{A.3} 
    	\includegraphics[width=0.8\hsize]{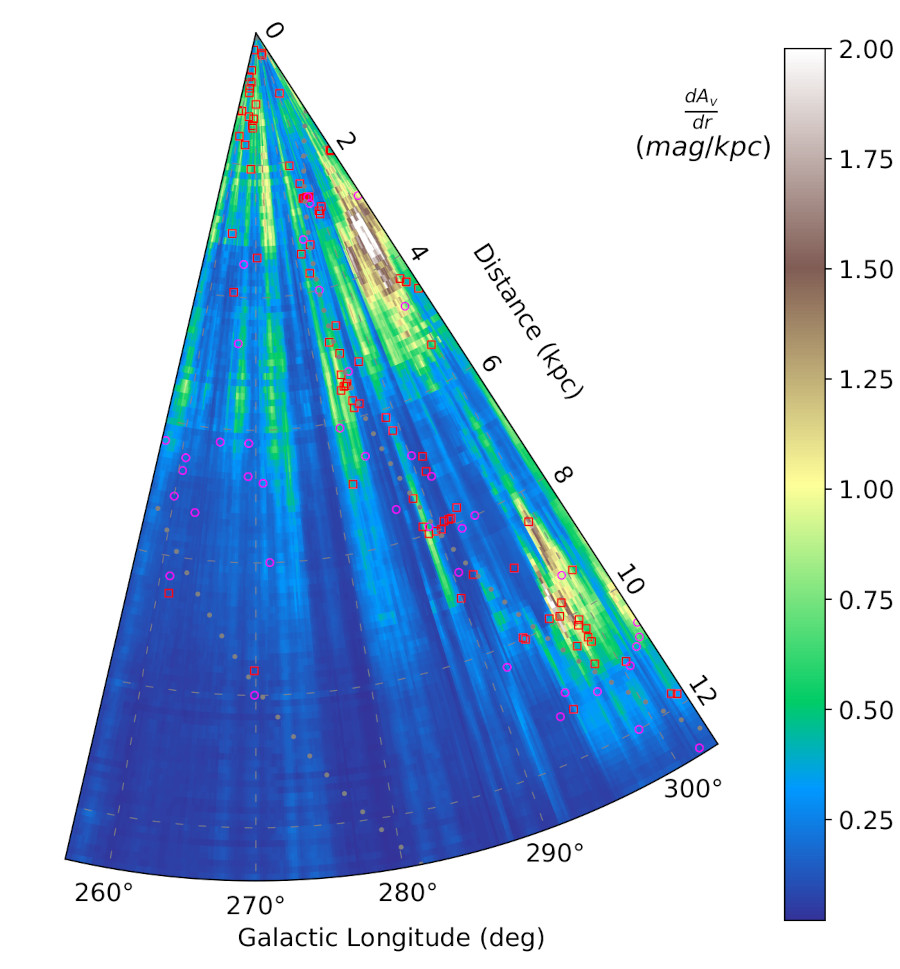}
        \caption{Standard deviation map of the Galactic Plane averaged over $|b|<1$\,deg in polar coordinates in the Carina-arm region. The standard deviations were computed from the MC model dropout prediction. The symbols are the same as in Fig.~\ref{multi_los_2mass_polar_sky}.}
    	\label{fig_std_polar_map}
\end{figure}

Figure~\ref{fig_std_polar_map} shows a face-on view of the GP in polar coordinates, obtained by averaging the 3D standard deviation map over the $|b|<1$\,deg range. Structures identified as probable artifacts coincide with high uncertainty predictions. Still, in Fig.~\ref{fig_std_polar_map}, high standard deviations are not always related to a false detection, but also to large uncertainties in the extinction value or to the merging of nearby structures.

\end{appendix}

\end{document}